\documentstyle[prd,aps,floats]{revtex}

\begin{document}
\draft


\title{Brans-Dicke brane cosmology} \author{Lu\'{\i}s E.
  Mendes$^\dagger$ and Anupam Mazumdar$^{*}$}
\address{$^\dagger$Astronomy Centre, University of Sussex, Falmer, BN1
  9QH, U.~K\\
  $^{*}$Astrophysics Group, Blackett Laboratory, Imperial College,
  London SW7 2BW, U.~K} \date{\today} \maketitle
\begin{abstract}
  A five dimensional brane cosmology with non-minimally coupled scalar
  field to gravity is considered in the Jordan frame. We
  derive the 
  effective four dimensional field equations on a $3+1$ dimensional
  brane where the fifth dimension is assumed to have an orbifold
  symmetry.  We show that energy conservation still holds for matter
  on the brane in spite of the existence of a scalar field (the
  Brans-Dicke field) in the bulk. Finally, we discuss
  some cosmological consequences of this setup.
 \end{abstract}



\section{Introduction}

Recently the possibility that the observable Universe is a brane-world
\cite{misha} embedded in a higher dimensional space-time has generated
a great deal of interest. Such a claim is motivated by the
strongly coupled sector of $E_8 \times E_8$ heterotic string theory
which can be described by a field theory living in $11$ dimensional
space-time \cite{horava}.  The $11$ dimensional world is comprised of
two $10$ dimensional hypersurfaces embedded on the fixed points of an
orbifold and the matter fields are assumed to be confined to these
hypersurfaces 
which in this scenario are known to be $9$ branes. After compactifying
the $11$ dimensional theory on a Calabi-Yau three fold, one obtains an
effective $5$ dimensional theory \cite{lukas} which has a structure
of two $3$ branes situated on the orbifold boundaries. The theory
allows $N=1$ supergravity with gauge and chiral multiplets on the two
$3$ branes. Motivated by the above setup there has been an attempt to
understand the case where the bulk is $5$ dimensional anti de-Sitter
space. It was shown that gravity could also be localized in this
scenario \cite{lisa}. These authors demonstrated that in a background of
a special non-factorizable geometry an exponential warp factor
multiplies the Poincar\'e invariant 3+1 dimensions in the metric. The
model consists of two $3$ branes situated rigidly along the 5th
dimension compactified on a $S^{1}/Z_2$ orbifold symmetry.  In order
to realize this simple scenario the two branes must 
have opposite tensions.

It was realized that the brane-world construction could modify the
early Universe considerably and, furthermore, the two branes with
opposite tensions should behave differently \cite{jing}.  The most
important observation was the departure of the evolution equation on
the brane from the standard four dimensional evolution equation when
no branes are present~\cite{bin}. The presence of branes and the
requirement that the fields should be localized lead to a
non-conventional cosmology which requires a more exhaustive study.
Some attention has been devoted to the effective gravity induced on
the brane~\cite{brane} and a great emphasis was placed on
inflationary cosmology~\cite{numero} and, more recently,
post-inflationary brane cosmology has been considered in Ref.~\cite{anu}. 
Apart from inflationary
and post-inflationary brane cosmology, there has been a great deal of
interest in the four dimensional cosmological constant problem
\cite{nim,kachru,bin1}. These authors have considered a five
dimensional action with a scalar field non-minimally coupled
to the five dimensional gravity and to the four dimensional brane
tension. There has also been some discussion on the localization of
gravity in this setup \cite{gomez}.  In this paper we would like to
extend the calculation made in Ref.~\cite{bin} to include the dynamics
of a scalar field which lives on the brane and the bulk. We then
study the effective field equations on the $3+1$ dimensional brane
locally and globally in presence of the other brane which is assumed
to be rigidly located on the orbifold symmetry along the 5th
dimension. Finally, we discuss the pertaining cosmology.
 

\section{Brans-Dicke brane model in $5$ dimensions}

In this article we restrict ourselves to one extra space-like
dimension with a scalar field non-minimally coupled to gravity in $5$
dimensions. We analyze the physics in the Jordan frame where the
interpretation of the results is straightforward, as we shall see
in the next sections. We believe that such a scalar field in the five
dimensional theory can be a dilaton which is purely an outcome of
dimensional reduction  
from some higher dimensional theory to $5$ dimensional
space-time~\footnote{This might not be the case if gravity had to be
  localized in a particular brane (see Ref.~\cite{kelly}).}.  We
assume the following action for the rest of our discussion
\begin{equation}
\label{action}
S_{5}=-\frac{1}{2 \kappa_{(5)}^2}\int d^5x \sqrt{-\tilde
  g}\left(\phi\tilde R -    
\frac{\omega}{\phi}\partial_{\rm A}\phi \partial ^{\rm A} \phi \right) 
+ \int d^{5}x \sqrt{-\tilde g} {\cal L}_{\rm m}\,,
\end{equation} 
where $\tilde R$ is the Ricci scalar associated with the 5-dimensional
space-time metric $\tilde g_{\rm AB}$, $\phi$ is a scalar field which we
will call Brans-Dicke (BD) field, $\omega$ is a dimensionless
coupling constant which determines the coupling between gravity and
the BD scalar field and finally, ${\cal L}_{\rm m}$ represents the
Lagrangian for the matter fields.  Latin indices denote $5$
dimensional components (${\rm A,B}=0,\ldots,5$) and for our
convenience we choose $\kappa_{5}^2=1$. The variational derivative of
the action Eq.~(\ref{action}) with respect to $g_{\rm AB}$ and $\phi$
yields the field equations~\cite{wein}
\begin{eqnarray}
\label{eqm0}
\tilde R_{\rm AB}- \frac{1}{2} \tilde g_{\rm AB} \tilde R  & = &
\frac{\omega}{\phi^2} 
\left[\phi_{;\rm A}\phi_{; \rm B} 
-\frac{1}{2} \tilde g_{\rm AB}\phi_{; \rm C}\phi^{;\rm C}\right] 
+\frac{1}{\phi}\left[\phi_{;\rm AB}-\tilde g_{\rm AB} {\phi^{;\rm C}}_{;\rm C} 
  \right] + \frac{\tilde T_{\rm AB}}{\phi} \, \, , \\
\label{eqm1}
\Box \phi  & = & \frac{\tilde T}{3\omega +4}\,,
\end{eqnarray}
where ${\tilde T=\tilde T^{\rm C}}_{\rm C}$ is the trace of the
energy-momentum tensor of the matter content of $5$ dimensional
space-time. Notice the $3\omega+4$ denominator in the right hand side
of the BD field equation instead of the familiar $2\omega+3$ in the
$4$-dimensional case~\cite{barrow}. This is determined by requiring
the validity of the equivalence principle in our setup (see Weinberg's 
book~\cite{wein} for a discussion of this topic in the context of
$4$-dimensional Brans-Dicke theory).

Before we discuss the energy-momentum tensor, let us
define the five dimensional metric which has the following form
\begin{eqnarray}
\label{met0}
ds^2=\tilde g_{\rm AB}dx^{\rm A}dx^{\rm B}=g_{\mu
  \nu}(x^{\mu},y)\, dx^{\mu}dx^{\nu} + b^2(x^{\mu},y) \, dy^2 \, ,
\end{eqnarray} 
where $\mu, \nu =0,\ldots,3$, and $y$ is the coordinate associated
with the fifth dimension which is
assumed to be compact with a range $-1/2 \leq y\leq 1/2$. We also
assume an orbifold symmetry along the fifth direction $y \rightarrow
-y$. As we shall see in the coming sections, this will help us to
simplify our calculation. Since the opposite points along the fifth
dimension are identified, we will only be interested in the interval
$0\leq y \leq 1/2$, which is the spacing between the two branes
situated at $y=0$ and $y=1/2$. Next we define the energy-momentum
tensor
\begin{eqnarray}
\label{em0}
\tilde {T^{\rm A}}_{\rm B}= {T^{\rm A}}_{\rm B}\arrowvert_{\rm bulk} +
{T^{\rm A}}_{\rm B} \arrowvert_{\rm brane}\,,
\end{eqnarray}
where the subscripts brane and bulk suggest the corresponding
energy-momentum tensors. For simplicity we assume the bulk is devoid
of matter other than the BD scalar field. The brane matter fields are
stuck at $y=0$ and $y=1/2$ with the following energy
momentum tensor
\begin{eqnarray}
\label{em1}
T^{\rm A}_{\rm B}\arrowvert_{\rm brane}&=& \frac{\delta(y)}{b}{\rm
  diag}(-\rho,p,p,p,0)\,, \\
\label{em2} 
T^{\rm A}_{\rm B}\arrowvert_{\rm brane *}&=&
\frac{\delta(y-1/2)}{b}{\rm diag} 
(-\rho_{*},p_{*},p_{*},p_{*},0)\,,
\end{eqnarray}
where the subscript `$*$' corresponds to the matter on the brane at
$y=1/2$. While writing the above expressions we have taken the
infinitely thin brane limit.

\section{Equations of motion}
Since we are interested in exploring the flat cosmology, we consider a $5$
dimensional flat metric ansatz of the following form
\begin{eqnarray}
\label{met1}
ds^2=-n^2(\tau,y)d\tau^2+a^{2}(\tau,y)\delta_{ij}dx^{i}dx^{j} +
b^2(\tau,y) dy^2\,,
\end{eqnarray}
where $i,j =1,2,3$. With this metric ansatz we are now able to write the 
equations of motion. The $(0,0)$ component reads
\begin{eqnarray}
\label{eq:00}
3 \left[\frac{\dot a}{a}\left(\frac{\dot a}{a}+\frac{\dot b}{b}\right)
 - \frac{n^2}{b^2} \left(\frac{a^{\prime
        \prime}}{a}+\frac{a^{\prime}}{a}\left(\frac{a^{\prime}}{a}- 
\frac{b^{\prime}}{b}\right)\right)\right] = \hspace{5cm} & & \nonumber \\
 -\frac{\dot{\phi}}{\phi}\left(3\frac{\dot{a}}{a} + \frac{\dot{b}}{b}
-\frac{\omega}{2} \frac{\dot{\phi}}{\phi}\right) + \left( \frac{n}{b}
\right)^2 \left[ \frac{\phi^{\prime\prime}}{\phi} +
  \frac{\phi^{\prime}}{\phi} \left( 3 \frac{a^{\prime}}{a} -
    \frac{b^{\prime}}{b} + \frac{\omega}{2} \frac{\phi^{\prime}}{\phi}
  \right)\right] + \frac{\tilde T_{00}}{\phi} \, \, ,
\end{eqnarray}
the $(i,j)$ component is given by
\begin{eqnarray}
\label{eq:ij}
\left\{-2\frac{\ddot a}{a} -\frac{\ddot b}{b} + \left[\frac{\dot a}{a}
    \left(-\frac{\dot 
      a}{a}+2\frac{\dot n}{n} 
\right)+ \frac{\dot b}{b} \left(-2\frac{\dot
    a}{a}+\frac{\dot n}{n} 
\right)\right]+ 
\left(\frac{n}{b}\right)^2 \left[2\frac{a^{\prime \prime}}{a}+\frac{n^{\prime
      \prime}}{n} + \frac{a^{\prime}}{a} 
  \left(\frac{a^{\prime}}{a}+2\frac{ 
n^{\prime}}{n}\right)-\frac{b^{\prime}}{b}
\left(\frac{n^{\prime}}{n}+2\frac{a^{\prime}}
{a}\right)
\right]\right\} \delta_{ij}  & =  & \nonumber \\
\left\{ \frac{\ddot{\phi}}{\phi} + \frac{\dot{\phi}}{\phi} \left(2
    \frac{\dot{a}}{a} + \frac{\dot{b}}{b}- \frac{\dot{n}}{n} +
    \frac{\omega}{2} \frac{\dot{\phi}}{\phi} \right) -
  \left(\frac{n}{b}\right)^2 \left[ \frac{\phi^{\prime\prime}}{\phi} +
    \frac{\phi^{\prime}}{\phi} \left(2 
    \frac{a^{\prime}}{a}  \frac{b^{\prime}}{b}+ \frac{n^{\prime}}{n} +
    \frac{\omega}{2} \frac{\phi^{\prime}}{\phi} \right)\right]
  \right\} \delta_{ij} + \left( \frac{n}{a}\right)^2
  \frac{\tilde T_{ij}}{\phi} \, \, , 
\end{eqnarray}
the $(0,5)$ component takes the form
\begin{eqnarray}
\label{eq:05}
3 \left(\frac{\dot{a}}{a}\frac{n^{\prime}}{n} + \frac{\dot{b}}{b}
  \frac{a^{\prime}}{a} -
  \frac{\dot{a}^{\prime}}{a}\right) = 
\frac{\dot{\phi}^{\prime}}{\phi} - \frac{\dot{\phi}}{\phi}\left(
  \frac{n^{\prime}}{n} - \omega \frac{\phi^{\prime}}{\phi}
\right) - \frac{\dot{b}}{b} \frac{\phi^{\prime}}{\phi} \, \, ,
\end{eqnarray}
and, finally, for the $(5,5)$ component one has
\begin{eqnarray}
\label{eq:55}
3 \left[ -\left(\frac{\ddot a}{a} + \frac{\dot a}{a}\left(\frac{\dot
        a}{a} - \frac{\dot n}{n}\right)\right) + \left( \frac{n}{b} \right)^2
  \left( \frac{a^{\prime}}{a}\left(\frac{a^{\prime}}{a} + 
    \frac{n^{\prime}}{n}\right) \right) \right] = \hspace{5cm} & & \nonumber \\
\frac{\ddot{\phi}}{\phi} +\frac{\dot{\phi}}{\phi} \left( 3
  \frac{\dot{a}}{a} - \frac{\dot{n}}{n} + \frac{\omega}{2}
  \frac{\dot{\phi}}{\phi} \right) - \left( \frac{n}{b}\right)^2 
  \frac{\phi^{\prime}}{\phi} \left( 3\frac{a^{\prime}}{a}  +
    \frac{n^{\prime}}{n} - \frac{\omega}{2}
  \frac{\phi^{\prime}}{\phi} \right) + \left( \frac{n}{b}\right)^2
\frac{\tilde T_{55}}{\phi} \, \, .
\end{eqnarray}
The equation of motion for the BD field reads
\begin{eqnarray}
\label{eq:bdfield}
\ddot{\phi} + \dot{\phi} \left( 3 \frac{\dot{a}}{a} +
  \frac{\dot{b}}{b} - \frac{\dot{n}}{n}\right) - \left( \frac{n}{b}
\right)^2 \left[ \phi^{\prime\prime} + \phi^{\prime} \left( 3
    \frac{a^{\prime}}{a} - \frac{b^{\prime}}{b} + \frac{n^{\prime}}{n}
  \right)\right] & = & - n^2 \frac{\tilde T}{3 \omega+4} 
\end{eqnarray}
where the dot corresponds to the time derivative with respect to $\tau$
and the prime corresponds to derivatives with respect to $y$. We make
the assumption that the metric and the BD field are continuous across
the branes localized at $y=0$ and $y=1/2$. However, their derivatives
can be discontinuous at the brane positions in the $y$ direction. This
suggests the second derivatives of the scale factor and the BD
field will have a Dirac delta function associated with the positions
of the branes. Since the matter is localized on the branes it will
introduce a delta function in the Einstein equations which will be
matched by the distributional part of the second derivatives of the
scale factor and BD field. For instance at $y=0$, we have
\begin{eqnarray}
\label{cond}
f^{\prime \prime}=\widehat{f^{\prime \prime}}+[f^{\prime}]\,\delta (y)\,,
\end{eqnarray}
where the hat marks the non-distributional part of the second
derivative of the quantity. The part associated with a delta
function, $[f']$, is a jump in the derivative of $f$. Here $f$ could
be any of the three quantities $a$, $n$ or $\phi$. The jump in $f$ at
$y=0$ can be written as
\begin{eqnarray}
[f^{\prime}]=f^{\prime}(0^{+})-f^{\prime}(0^{-})\,,
\end{eqnarray}
and the mean value of the function $f$ at $y=0$ is defined by
\begin{eqnarray}
\label{eq:meandef}
\sharp f \sharp = \frac{f(0^{+})+f(0^{-})}{2}\,.
\end{eqnarray}
After substituting Eq.~(\ref{cond}) in the Einstein equations it is
possible to find out the jump conditions for $a$ and $n$ by matching
the Dirac delta functions appearing in the left-hand side of the
Einstein equations to the ones coming from the energy-momentum tensor
Eqs.~(\ref{em0}--\ref{em2}).  For the BD field one has to use
Eq.~(\ref{eq:bdfield}) to evaluate the jump condition.  We get the
following results for the jump conditions
\begin{eqnarray}
\label{eq:jumpa}
\frac{[a^{\prime}]_0}{a_0 b_0} & = & -\frac{1}{(3\omega
  +4)\phi_0}\left(  p+(\omega +1) \rho \right) \,, \\
\label{eq:jumpn}
\frac{[n^{\prime}]_0}{n_0b_0}&=&\frac{1}{(3\omega
  +4)\phi_0}\left(3(\omega +1)p+(2 \omega +3)\rho\right)\,, \\
\label{eq:jumpbd} 
\frac{[\phi^{\prime}]_0}{\phi_0 b_0} & = & \frac{1}{(3\omega
  +4)\phi_0}\left(3p-\rho \right)\, ,
\end{eqnarray} 
where the subscript `$0$' stands for the brane at $y=0$. Similar
conditions hold for the brane fixed at $y=1/2$ by replacing `$0$' by
`$1/2$' and $\rho$ and $p$
by $\rho_*$ and $p_*$. The first two conditions, Eqs.~(\ref{eq:jumpa})
and (\ref{eq:jumpn}), are equivalent to Israel's junction conditions in
general relativity~\cite{israel} (see Ref.~\cite{bin} for a discussion of
its application in the context of brane world). It is important to
note that the above jump conditions at $y=0$ depend on the energy
density and the pressure component of the brane world. Remarkably for
the radiation dominated phase on the brane, $\rho=3p$, the jump
condition for $\phi$ vanishes suggesting  that the BD field takes a
constant value on the bulk and the brane. This has important  
consequences which we shall discuss later on. 

Taking the jump of the ($0,5$) component of the Einstein equation and
substituting Eqs.~(\ref{eq:jumpa}-\ref{eq:jumpn}) we get the
continuity equation for the matter on the brane
\begin{eqnarray}
  \label{eq:cons}
  \dot \rho +3 \left( \rho + p \right) \frac{\dot a}{a}
= 0 \,.
\end{eqnarray} 
The fact that the energy content of the brane is still conserved in
this scenario seems to be at odds with recent results obtained
independently by several authors who conclude that the
presence of a dilaton in the bulk will lead to a non-trivial coupling
with the matter on the brane which from the point of view of an
observer living on the brane would be seen as matter leaking from the
brane~\cite{bin1,wands,mennim,mart}. The reason why our situation is
different is because the coupling between the BD field and ordinary
matter on the brane was chosen in order to satisfy the equivalence
principle, as was pointed out above. Had we chosen the coupling
in~(\ref{eq:bdfield}) to be the usual $4$-dimensional value
$(2\omega+3)^{-1}$ we would end up with a situation where the
conservation equation~(\ref{eq:cons}) would not hold and energy could
leak from the brane. 

Taking the mean value (in the sense of Eq.~(\ref{eq:meandef})) of the
$(5,5)$ component of Einstein's equations we can now obtain a
Friedmann type equation 
on the brane by following a very similar procedure to the one
introduced in~\cite{bin}. Using the fact that due to the orbifold
symmetry $y \leftrightarrow -y$, $\sharp f^{\prime}\sharp =0$ (for $f$
any of the quantities $a$, $n$ or $\phi$), we can discard all the
terms involving mean values in the average of the $(5,5)$ component of
the Einstein equations.  The equation so obtained will still involve a
term containing $\ddot{\phi}$, but we can use the mean value of the BD
field equation~(\ref{eq:bdfield}) to write this in terms of $a$, $b$
and $n$, and their derivatives. After a tedious calculation we obtain
the Friedmann type equation on the brane
\begin{eqnarray}
  \label{eq:friedtype}
  \frac{\ddot a_0}{a_0}+\left(\frac{\dot a_0}{a_0}\right)^2+
  \frac{\omega}{6}\left(\frac{\dot \phi_0}{\phi_0}\right)^2 = 
  -\frac{1}{24 (3\omega +4)\phi_0^2}
  \left[ \left( 2\omega+3 \right) \rho^2 + 6\left( \omega+1 \right)
    p\rho +3 p^2 + \frac{2\omega\left( 3 p -\rho\right)^2}{3\omega+4
      }\right] \,\, .
\end{eqnarray}
While deriving the above equation we have assumed the extra dimension
to be static, $b=b_0$, from the point of view of the brane observer and
we have also fixed time in such a way that $n_0=1$. This corresponds
to the usual choice of time in conventional cosmology. It is
interesting to note that by taking the limit $\omega \rightarrow
\infty$ in the right-hand side of Eq.~(\ref{eq:friedtype}), we obtain
exactly the same expression as in general relativity (compare with Eq.
(20) in Ref.~\cite{bin}). However, the above expression differs from
the corresponding equation of motion without the presence of branes
(see Ref.~\cite{barrow}), namely due to the presence of quadratic
terms in $\rho$ and $p$ and the absence of linear terms in the energy
density and pressure in~(\ref{eq:friedtype}).

From the mean value of the BD field equation we obtain an equation of
motion for $\phi$ on the brane
\begin{eqnarray}
  \label{eq:bdeq}
  \ddot \phi_0 +3\frac{\dot a_0}{a_0} \dot{\phi_0}=
  \frac{\omega \left(\rho -3p\right)^2}{4(3\omega +4)^2  
    \phi_0} \,\, .
\end{eqnarray}
Note that in order to obtain Eq.~(\ref{eq:bdeq}) we also have to
assume the non-distributional part of $\phi^{\prime\prime}$ vanishes,
otherwise, a term involving $\widehat{\phi^{\prime\prime}}$ will
appear in the BD field equation. As we shall see in the next Section
it is possible to obtain cosmologically interesting solutions which
verify this condition. Together with the conservation
equation~(\ref{eq:cons}), Eqs.~(\ref{eq:friedtype})
and~(\ref{eq:bdeq}) determine the cosmology of the brane at $y=0$.
Comparing equation~(\ref{eq:bdeq}) with the usual BD field equation in
$4$ dimensional BD theory with no branes we note there is an extra
power of $\rho -3p$ and an extra factor of $\phi_0^{-1}$ on the right
hand side of the BD field equation in the brane model.
\section{Topological constraints}

As in the general relativity case, in our setup matter in one
brane is also constrained by matter on the second brane (see
Ref.~\cite{bin} for a detailed discussion of this point). In order to
get the constraints, we proceed as in Ref.~\cite{bin}. We will use the
following ansatz for the solution
\begin{equation}
  \label{eq:ansatz_a}
  a = a_0(t) + \left( \frac{|y|}{2}-\frac{y^2}{2}\right)
  \left[a^{\prime}\right]_0 -\frac{y^2}{2}
  \left[a^{\prime}\right]_{1/2 }\,,
\end{equation}
where the subscripts `$0$' and `$1/2$' denote the branes at $y=0$ and
$y=1/2$ with a similar expression for $n$.  As to the BD field $\phi$,
we assume it has the following form
\begin{equation}
  \label{eq:ansatz_bd}
  \phi = \phi_0(t) + \frac{\left[\phi^{\prime}\right]_0}{2} |y| \, . 
\end{equation}
Notice that the term quadratic in $y$ which is present in $a$ and $n$
is absent from $\phi$. This difference will be analyzed below.  Since
$n_0(t)$ is a completely arbitrary function which can be fixed by a
choice of time coordinate we fix $n_{0}(t)=1$ in order to get the
conventional definition of time. For $b$ we use
\begin{equation}
  \label{eq:ansatz_b}
  b = b_0 \, \, \, ,
\end{equation}
where $b_0$ is a constant. This ansatz corresponds to stabilizing the
fifth dimension.  This is by no means a trivial assumption but for
simplicity we assume the fifth dimension to be static.
Using the above ansatz for the static extra dimension the  
$(0,0)$ component of the Einstein equations for the brane situated at
$ y=0$ yields 
\begin{equation}
  \label{eq:00_ansatz}
  \left( \frac{\dot a_0}{a_0} \right)^2 + \frac{\dot \phi_0}{\phi_0}
  \left( \frac{\dot a_0}{a_0} -\frac{\omega}{6}
    \frac{\dot\phi_0}{\phi_0} \right) = \frac{1}{b_0^2} \left[
    \frac{1}{4} \left( \frac{[a^{\prime}]_0}{a_0}\right)^2 +
    \frac{\omega}{24} \left( \frac{[\phi^{\prime}]_0}{\phi_0}\right)^2
   + \frac{1}{4} \frac{[a^{\prime}]_0}{a_0}
   \frac{[\phi^{\prime}]_0}{\phi_0} - \frac{[a^{\prime}]_{1/2}}{a_0} -
 \frac{[a^{\prime}]_0}{a_0}\right]\, \, .  
\end{equation}
In order to be consistent with Eq.~(\ref{eq:friedtype}) the right-hand
side of this equation must be quadratic in the energy density.
Noticing that the jump conditions for $a$ and $n$ are linear in the
energy density, this means that the right-hand side of
Eq.~(\ref{eq:00_ansatz}) must contain only terms quadratic in the jump
for the various quantities involved. Terms linear in the
jump must therefore vanish yielding the condition
\begin{equation}
  \label{eq:linear_jump_(00)}
  [a^{\prime}]_0 =- [a^{\prime}]_{1/2}\, \, .
\end{equation}

Following a similar procedure with the $(i,j)$ components of the
Einstein equations, with the help of the previous constraint
Eq.~(\ref{eq:linear_jump_(00)}) we obtain our last constraint
\begin{equation}
  \label{eq:linear_jump_(ij)}
  [n^{\prime}]_0 = - [n^{\prime}]_{1/2} \, \, .
\end{equation}
These constraints have the same form in terms of the jumps in $a$ and
$n$ as in the general relativity case. However, using the jump
conditions Eqs.~(\ref{eq:jumpa}) and~(\ref{eq:jumpn}) they result in
different relations between the energy density and pressure in both
branes than the ones found in Ref.~\cite{bin}.

Assuming the matter on the branes obeys an equation of state
$p=\epsilon\rho$ for the brane at $y=0$ and $p_* =\epsilon_{*}\rho_*$
for the brane at $y=1/2$, Eqs.~(\ref{eq:linear_jump_(00)})
and~(\ref{eq:linear_jump_(ij)}) read
\begin{eqnarray}
  \label{eq:constraints_pen1}
  \left[ \epsilon+ \left( \omega +1\right)\right] \rho
  \frac{a_0}{\phi_0}& = & 
  -\left[ \epsilon_{*}+\left(\omega +1\right)\right] \rho_{*}
  \frac{a_{1/2}}{\phi_{1/2}}\,,
\end{eqnarray}
and,
\begin{eqnarray}
  \label{eq:constraints_pen2}
  \left[3(\omega +1)\epsilon+ 
    (2 \omega +3)\right] \rho \frac{n_0}{\phi_0} & = &
  -\left[3(\omega +1)\epsilon_*+ 
    (2 \omega +3)\right] \rho_* \frac{n_{1/2}}{\phi_{1/2}}\, \, .
\end{eqnarray}

Had we introduced a quadratic term in $y$ in the ansatz for the BD
field in Eq.~(\ref{eq:ansatz_bd}), we would end up with a third
constraint
\begin{equation}
  \label{eq:linear_jump_bd}
  [\phi^{\prime}]_0 = - [\phi^{\prime}]_{1/2} \, \, .
\end{equation}
We must note that this constraint is formally equivalent
to~(\ref{eq:linear_jump_(00)}) and~(\ref{eq:linear_jump_(ij)}) in the
sense that the three constraints require the jump in the first
derivative of a given quantity across one brane to have the same
absolute value but an opposite sign 
as the same jump across the second brane. Since
we have only two quantities to fix, namely $\rho_{*}$ and $p_{*}$,
this extra relation would possibly be verified only for a discrete set of
values of $\omega$ and/or $\epsilon$. Therefore by discarding the
quadratic term in $\phi$ we are automatically free from this problem.
In any case our ansatz can always be interpreted as a power series expansion in
$y$ of a more complex solution in which case it is valid for $y$ small
enough. Since we are mainly interested in the cosmology of the brane
at $y=0$ we are well within the domain where it is safe to consider
only the lower order terms in the power series.

Although we are not going to discuss further the problem of
constraints in this paper, we suggest that a much more elegant
solution could possibly be found with a varying $\omega$ in the bulk.
This could in principle be implemented in the context of the so called
scalar-tensor theories where $\omega=\omega(\phi)$. The third
constraint would then fix the values of $\omega$ at $y=1/2$ as
a function of $\omega$ on the brane at $y=0$, therefore partially
constraining the functional form of $\omega=\omega(\phi)$. Another possibility
which would make the constraints unnecessary is a model with a single
brane. In that case,  Eq.~(\ref{eq:friedtype}) would hold without
any further ado. We must point out however that the presence of a
second brane is favored for several reasons
(see~Refs.\cite{lukas,kelly}). As a final remark, let us note that
constraints of this type are expected to appear in models with compact
extra dimensions~\cite{bin}.


\section{Exact cosmological solutions}

We will now obtain some simple solutions for $a$ and $\phi$ which will
allow us to discuss cosmology in our setup. For what follows we will
discard the part of our ansatz in Eq.~(\ref{eq:ansatz_a}) (and
similarly for $n$) which is proportional to $y^2$. In this context
this means that we will no longer have to worry about the constraints
discussed in the previous section.  As before we assume the extra
space like dimension is stabilized ($b$ is constant) and choose time
such that $n_0=1$.

From Eq.~(\ref{eq:00_ansatz})
and using the jump conditions Eqs.~(\ref{eq:jumpa}--\ref{eq:jumpbd})
we obtain the Friedmann equation which reads
\begin{equation}
  \label{eq:friedeq}
  \left( \frac{\dot{a}_0}{a_0}\right)^2 + \frac{\dot{\phi}_0}{\phi_0}
  \left( \frac{\dot{a}_0}{a_0} - \frac{\omega}{6}
    \frac{\dot{\phi}_0}{\phi_0} \right) = \frac{1}
      {4 \left( 3 \omega + 4 \right)^2 \phi_0^2} \left[
    \frac{\omega}{6} \left( 3 p -\rho\right)^2 + \left
      (2+3\,\omega+{\omega}^{2}\right ){\rho}^{2}-p\rho\,\omega-2\,{p}^{2} 
\right] \, \, .  
\end{equation}
Using our ansatz in the BD field equation, we obtain again
Eq.~(\ref{eq:bdeq}), as we should expect. Together with the
conservation equation ~(\ref{eq:cons}), Eqs.~(\ref{eq:bdeq})
and~(\ref{eq:friedeq}) form a complete system which will allow us to
study the cosmology on the brane. A careful look at the above
mentioned equations suggests that once the equation of state on the
brane has been specified, it is possible to obtain a power law
solution to the scale factor, BD field and the matter content on the
brane as a function of time.

Here we consider only the particular cases of radiation ($p=\rho/3$)
and dust ($p=0$) as well as the vacuum energy case ($p=-\rho$).  For a
radiation dominated Universe we obtain $\phi_{0} = {\rm const.}$ from
Eq.~(\ref{eq:bdeq}), which also holds for the usual BD solution in
four dimensions. It is noticeable that in this case the solution is
exactly the same as in the case of general relativity (see
Ref.\cite{bin}):
\begin{equation}
  \label{eq:rad}
  a_{0} \propto \tau ^{1/4}\, ;  ~~\rho_{0} \propto a_{0}^{-4}\, ; 
   ~~\phi_{0} = {\rm const}\,. 
\end{equation}

For the matter dominated era the solution can be obtained from
Eqs.~(\ref{eq:cons}), (\ref{eq:bdeq}) and (\ref{eq:friedeq}), and is 
given by
\begin{equation}
  \label{eq:dust}
  a_{0} \propto \tau
  ^{\frac{3\omega+3+\sqrt{3}\sqrt{2\omega+3}}{9\omega+12}} \, ; 
~~\rho_{0} \propto a_{0}^{-3}\, ;
~~\phi_{0} \propto \tau ^{\frac{1+\sqrt{3}\sqrt{2\omega+3}}{3\omega+4}}\,,
\end{equation}
where we have taken the growing mode for the scale factor and we have
restricted to large $\omega$ by truncating terms ${\cal O}(1/\omega^2)$. It
is important to notice that for large $\omega$ the solutions approach
the brane general relativity case $a\propto\tau^{1/3}$. If the
solution for the scale factor in the $4$ dimensional BD case is
written as $a\propto\tau^\alpha$ then in our model the solution takes
the form $a\propto\tau^{\alpha/2+{\cal O}(1/\sqrt{\omega})}$. In our
situation however the expansion during the dust era is faster compared
with that in the general relativity case. This will have some
implications during the rapid oscillations of the scalar 
field in our brane after inflation and during preheating when the
expansion mimics that of a pressureless fluid.

It is important to mention here that our results are valid only prior
to nucleosynthesis and at late times the Friedmann equation 
should converge to the standard case with $H \propto  \sqrt {\rho}$ in 
order to satisfy the stringent constraints coming from big bang
nucleosynthesis. For this to happen one may need to take the brane tension into
account~\cite{bin}. Denoting the brane tension by $\rho_{\Lambda}$, 
the total energy density in the brane will be given by $\rho_{\rm
  tot}=\rho_{\Lambda}+ \rho$ and the square of this quantity will
include the term $2\rho_{\Lambda}\rho$ linear in $\rho$. For
$\rho_{\Lambda} > 1 {\rm MeV}^4$ one can ensure compatibility with
nucleosynthesis constraints. In this way for 
energy scales larger than the brane tension we have the non-standard
behaviour, while as the universe expands and cools down we recover the 
standard behaviour. 

Finally, we discuss the cosmology of the vacuum energy dominated
stage. From the energy conservation equation we see that false vacuum
inflation ($\rho=$constant) is possible in this case for large enough
values of $\omega$. The scale factor and BD field are given by  
\begin{equation}
a_{0}(\tau) \propto \tau^{\frac{3\omega}{16}} \;
\phi_{0}(\tau)\propto\tau \; .
\end{equation}
The rate of expansion during inflation in the brane is
therefore slower when compared with the $4$ dimensional BD case
$a\propto\tau^{\omega + 1/2}$. As for the BD field, in the $4$
dimensional case $\phi\propto \tau^2$ and the evolution is also slower 
in the brane model.

Note that if the energy conservation equation~(\ref{eq:cons}) was not
satisfied  as in the models studied in~\cite{wands,mennim,mart} then
false vacuum inflation would not be possible because due to matter
leaking from the brane, even for the case $p=-\rho$ it would  not possible to 
keep the energy density constant anymore.

An interesting possibility which we have not analyzed here is the case
where the dynamics in the brane is driven by the evolution of the fifth 
dimension. A more detailed study of the cosmology in the model described in
this paper, including an analysis of the primordial perturbation
spectra produced by inflation in our setup will be discussed elsewhere. 

\section{Conclusions}
We have derived the effective four
dimensional field equations for the brane cosmology in presence of the
BD field in five dimensions. In our scenario, the conservation
equation for the matter field stuck to the brane still holds in spite
of the BD field present in the bulk. This is due to the fact that our
calculation was done in the Jordan frame. Some global
solutions with two 
branes such that the fifth dimension has an orbifold symmetry were
discussed. Finally we obtained the modified Friedmann equation in the
brane and cosmological solutions in the BD brane setup were discussed.
\section*{Acknowledgments}  
L.E.M is supported by FCT (Portugal) under contract PRAXIS XXI
BPD/14163/97. A. M. is supported by INLAKS foundation. The authors
are thankful to Jim Lidsey for fruitful discussions. We are extremely
grateful to Hassan Firouzjahi and Christophe Grojean for pointing out  
an error in our calculation which has led to a complete revision 
of our paper.


\end{document}